\documentclass[conference]{IEEEtran}
\IEEEoverridecommandlockouts

\usepackage{cite}
\usepackage{amsmath,amssymb,amsfonts}
\usepackage{algorithmic}
\usepackage{graphicx}
\usepackage{textcomp}
\usepackage{xcolor}
\usepackage{multicol}
\usepackage{multirow}
\usepackage{booktabs}
\usepackage{float}
\usepackage{caption}
\usepackage{tabularx}
\def\BibTeX{{\rm B\kern-.05em{\sc i\kern-.025em b}\kern-.08em
    T\kern-.1667em\lower.7ex\hbox{E}\kern-.125emX}}
\begin{document}

\title{MDN: Mamba-Driven Dualstream Network For Medical Hyperspectral Image Segmentation\\
\thanks{This work was supported by the National Natural Science Foundation of China (Grant No. 62471182, Grant No. 62101191), the foundation of Key Laboratory of Artificial Intelligence, Ministry of Education, P.R. China, and the Fundamental Research Funds for the Central Universities.}
}
\author{\IEEEauthorblockN{Shijie Lin}
\IEEEauthorblockA{\textit{Shanghai Key Laboratory of}\\ 
\textit{Multidimensional Information Processing} \\
\textit{East China Normal University}\\
Shanghai, China \\
71255904039@stu.ecnu.edu.cn}
\and
\IEEEauthorblockN{Boxiang Yun}
\IEEEauthorblockA{\textit{Shanghai Key Laboratory of}\\ 
\textit{Multidimensional Information Processing} \\
\textit{East China Normal University}\\
Shanghai, China \\
52265904012@stu.ecnu.edu.cn}
\and
\IEEEauthorblockN{Wei Shen}
\IEEEauthorblockA{\textit{Key Laboratory of Artificial Intelligence,}\\
\textit{Ministry of Education} \\
\textit{Shanghai Jiao Tong University}\\
Shanghai, China \\
wei.shen@sjtu.edu.cn}
\and
\IEEEauthorblockN{Qingli Li}
\IEEEauthorblockA{\textit{Shanghai Key Laboratory of}\\ 
\textit{Multidimensional Information Processing} \\
\textit{East China Normal University}\\
Shanghai, China \\
qlli@cs.ecnu.edu.cn}
\and
\IEEEauthorblockN{Anqiang Yang}
\IEEEauthorblockA{\textit{Department of Pathology} \\
\textit{Changning Maternal and Infant}\\
\textit{Health Hospital}\\
Shanghai, China \\
yaq1234@126.com}
\and
\IEEEauthorblockN{Yan Wang$^*$\thanks{$^*$Corresponding author.}}
\IEEEauthorblockA{\textit{Shanghai Key Laboratory of}\\ 
\textit{Multidimensional Information Processing} \\
\textit{East China Normal University}\\
Shanghai, China \\
ywang@cee.ecnu.edu.cn}
}

\maketitle
\begin{abstract}
Medical Hyperspectral Imaging (MHSI) offers potential for computational pathology and precision medicine. However, existing CNN and Transformer struggle to balance segmentation accuracy and speed due to high spatial-spectral dimensionality. In this study, we leverage Mamba's global context modeling to propose a dual-stream architecture for joint spatial-spectral feature extraction. To address the limitation of Mamba's unidirectional aggregation, we introduce a recurrent spectral sequence representation to capture low-redundancy global spectral features. Experiments on a public Multi-Dimensional Choledoch dataset and a private Cervical Cancer dataset show that our method outperforms state-of-the-art approaches in segmentation accuracy while minimizing resource usage and achieving the fastest inference speed. Our code will be available at https://github.com/DeepMed-Lab-ECNU/MDN.
\end{abstract}

\begin{IEEEkeywords}
Medical hyperspectral image segmentation, Mamba, dualstream.
\end{IEEEkeywords}

\section{Introduction}
\begin{figure}
 \vspace{-10pt}
    \centering
    \includegraphics[width=1\linewidth]{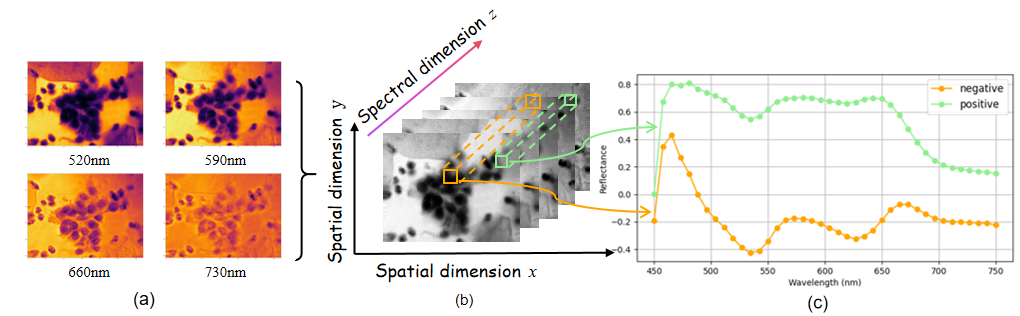}    \caption{(a): 3D cube of medical hyperspectral images of cervical cancer. (b): Morphology of cervical cancer cells at different wavelengths, (c): Spectral curve of cancer cells versus normal cells.}
    \label{fig:MHSI}
\vspace{-17pt}
\end{figure}

Medical Hyperspectral Imaging (MHSI), as an emerging imaging modality, captures the spectral differences between various tissues (Fig.\ref{fig:MHSI}(c)), by monitoring their responses to light across different wavelengths. This technology provides rich spectral and spatial information for medical image analysis   (Fig.\ref{fig:MHSI}(a) and  Fig.\ref{fig:MHSI}(b)). With advancements in AI, MHSI segmentation is crucial for distinguishing cancerous from non-cancerous regions. However, the high dimensionality and complexity of MHSI data pose significant challenges in extracting meaningful spectral-spatial features for accurate segmentation. 
 \par
Convolutional neural network (CNN)-based methods\cite{u-net, nnu-net, 3d-unet} are widely studied for their strong feature extraction capabilities in medical image segmentation. To handle MHSI’s high dimensionality, dual-channel networks integrating 1D CNNs \cite{dc-cnn,double-branch2020} or recurrent neural networks (RNNs)\cite{double-Bi-GRU, bengs2020spectralspatial} are used to extract spatial-spectral features. Other approaches,  like principal component analysis \cite{wang2021pca} or spectral low-rank decomposition \cite{yun2023factor} reduce spectral redundancy to enhance efficiency. However, CNNs struggle to capture long-range dependencies due to local receptive fields, and RNNs fail to capture complex spectral patterns in long sequences fully.    
  \par

\begin{figure*}[ht]
\vspace{-10pt}
        \centering
            \includegraphics[width=1\linewidth]{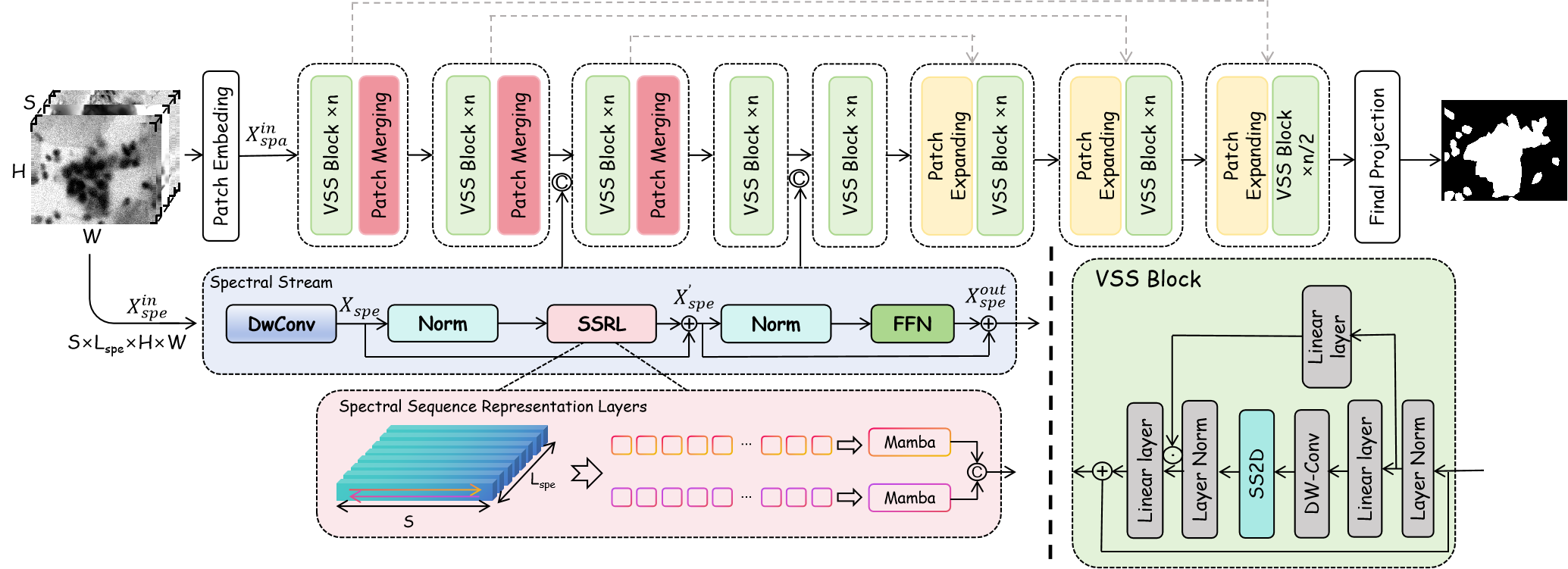}
                \caption{The proposed mamba-driven dualstream architecture. MHSI extracts spatial information and spectral information separately through the spatial stream and spectral stream. The spectral stream includes Depth-wise Convolution (DwConv),  Spectral Sequence Representation Layer (SSRL), and Feedforward Network (FFN).  }
        \label{FIG:2}
    \vspace{-10pt}
    \end{figure*}

Transformer-based methods leverage self-attention to model long-range dependencies and capture global contextual information. Some studies \cite{hong2021spectralformer, yun2021spectr} introduce Transformers to focus on spectral features, while others adopt a dual-stream architecture  \cite{sun2022spectral, yu2023dstrans, long2023dualswintr, tang2023double}, applying self-attention separately to spatial and spectral dimensions to capture both long-range dependencies.  These methods improve segmentation accuracy in complex imaging tasks. However, their quadratic complexity burdens hyperspectral data, and the lack of spectral redundancy handling complicates MHSI analysis.
\par 
Selective structured state-space models (Mamba) \cite{gu2023mamba} offer an efficient alternative to self-attention by modeling long-range dependencies with linear complexity. Mamba has been successfully applied to hyperspectral remote sensing tasks \cite{spectralmamba, s2mamba}, flattening central patches into spectral sequences to capture long-range spectral features and reduce spectral redundancy. However, Mamba's unidirectional aggregation and reliance on central spectral representation limit its ability to capture comprehensive global continuous spectral features.  
\par
To address (1) the speed-accuracy trade-off between CNNs and Transformers, and (2) the difficulty in extracting low-redundancy and globally continuous spectral features. This paper develops a novel dual-stream model. In the spectral stream, we integrate spatial information into the spectral dimension and focus on spectral feature extraction. We propose to use Mamba's parameter \(A\) to reduce spectral redundancy and flatten the feature maps into a recurrent spectral sequence to capture global spectral features. In the spatial stream, we enhance feature extraction using Vmunet \cite{ruan2024vmunet}. Finally, we combine spatial information with low-redundancy global spectral information via channel-wise concatenation, improving both segmentation accuracy and efficiency.  
 \par
 The main contributions of this paper are as follows:
\begin{enumerate}
    \item We develop a novel dual-stream model,  firstly applying Mamba to MHSI segmentation to capture long-range spatial relationships and global spectral features while reducing spectral redundancy. 
    \item We design a spectral stream that focuses on capturing spectral contextual information by flattening shallow feature maps into spectral feature sequences and recursively computing each spectral feature, effectively capturing the spectrum's contextual information.
    \item We evaluate our method on both public\cite{zhang2019multidimensional} and private datasets, showing competitive segmentation performance with lower computational overhead compared to other state-of-the-art methods.
\end{enumerate}

\begin{table*}[h]
\renewcommand\arraystretch{1}
\vspace{-10pt}
\setlength{\tabcolsep}{8pt}
\centering
\caption{Performance comparison with SOTA methods with Throughput(images/s) and computational complexity of MACs (G), the input size is \(320\times256\times60\) for MDC and \(256\times256\times40\) for HCC on both dataset. The best results are \textbf{highlighted}.}
\label{Tab:comparison_sota}
\begin{tabular}{c|l|cccc|cccc}
\toprule
\multicolumn{2}{c|}{\multirow{2}{*}{\textbf{Method}}} & \multicolumn{4}{c|}{\textbf{MDC}} & \multicolumn{4}{c}{\textbf{HCC}} \\ 
\cline{3-10}
\multicolumn{2}{c|}{} & \textbf{DSC}$\uparrow$ & \textbf{HD}$\downarrow$ & \textbf{Throughput}$\uparrow$ & \textbf{MACs(G)}$\downarrow$ & \textbf{DSC}$\uparrow$ & \textbf{HD}$\downarrow$ & \textbf{Throughput}$\uparrow$ & \textbf{MACs(G)}$\downarrow$ \\ 
\midrule

\multirow{4}{*}{\textbf{CNN}} 
& PCA-UNet \cite{wang2021pca} & 70.83 & 80.70 & 12.37 & \textbf{13.31} & 60.50 & 88.57 & 9.35 & \textbf{7.85} \\
& DualStream \cite{yun2023factor} & 74.12 & 78.32 & 10.41 & 250.89 & 73.52 & 73.18 & 20.74 & 201.20 \\
& 3D-UNet \cite{3d-unet} & 72.55 & 79.87 & 4.04 & 1110.77 & 64.74 & 83.20 & 5.18 & 1896.45 \\
& HyperNet \cite{hypernet} & 72.47 & 83.75 & 0.99 & 1512.24 & 75.75 & 71.15 & 6.23 & 810.37 \\ 
\midrule
\multirow{1}{*}{\textbf{RNN}} 
& CGRU-UNet \cite{bengs2020spectralspatial} & 73.14 & 82.49 & 7.50 & 35.98 & 73.26 & 76.69 & 7.32 & 26.64 \\ 
\midrule
\multirow{3}{*}{\textbf{Transformer}} 
& Transunet \cite{chen2021transunet} & 72.85 & 77.56 & 9.25 & 33.62 & 67.99 & 82.87 & 7.75 & 30.18 \\
& Swin-UNETR \cite{swinuneter} & 72.39 & 78.38 & 1.45 & 245.04 & 71.65 & 70.64 & 6.91 & 197.57 \\
& SpecTr \cite{yun2021spectr} & 73.66 & \textbf{76.92} & 1.40 & 1049.72 & 74.82 & 70.40 & 2.66 & 626.60 \\ 
\midrule
\multicolumn{2}{c|}{\emph{\textbf{ours}}} & \textbf{74.50} & 77.93 & \textbf{14.12} & 31.82 & \textbf{78.10} & \textbf{66.95} & \textbf{22.16} & 23.74 \\
\bottomrule
\end{tabular}
\vspace{-15pt}
\end{table*}

\section{METHOD}
\label{sec:format}
We focus on some basic concepts about State Space Models (SSM),  Structured State Space Sequence Model (S4) and Mamba before introducing our architecture.
\subsection{Preliminaries}
\subsubsection{SSM and S4}
Inspired by control system theory\cite{cellier2013continuous},  SSM is considered linear time-invariant systems that map a signal \(x(t)\in\mathbb{R} \) to  \(y(t)\in\mathbb{R}\) through the hidden state \(h(t)\in\mathbb{R}^{ N}\).  Mathematically, the procedure can be formulated as follows:  
\begin{equation}
\begin{aligned}
h'(t) &= Ah(t) + Bx(t), \\
y(t) &= Ch(t), 
\end{aligned}
\label{eq:ssm}
\end{equation}
where the parameters include \(A \in \mathbb{R}^{N \times N}\), \(B \in \mathbb{R}^{N \times 1}\) and \(C \in \mathbb{R}^{N \times 1} \) for a state size \(N\). Notably, \(A \) summarizes all past historical information. 
\par
To adapt continuous systems for deep learning tasks with discrete sequences, such as images and text,  S4\cite{s4} introduces a time-scale parameter \(\Delta\) to transform the continuous parameter \(A\) and \(B\) into discrete parameters \(\overline{A}\) and \(\overline{B}\).  For discrete sequences, Eq.~(\ref{eq:ssm}) can be expressed as below:   
\begin{equation}
\begin{aligned}
&\overline{A} = \exp (\Delta A), \\
&\overline{B} = (\Delta A)^{-1}(\exp(\Delta A) - I) \cdot \Delta B, \\
&h_k = \overline{A}h_{k-1} + \overline{B}x_k, \\
&y_k = Ch_k,
\end{aligned}
\label{eq:s4}
\end{equation}
where \(k\) denotes the time step. Each output \(y_k\) can be computed using a recurrent structure for fast inference or a convolutional structure for parallel training.
\par
However,  as noted in\cite{gu2021combining}, the matrices \(\overline{A}\), \(\overline{B}\) and \(\overline{C}\) share a crucial attribute. Regardless of the input sequence to S4, the values of  \(\overline{A}\), \(\overline{B}\) and \(\overline{C}\)  remain fixed, making it impossible to perform targeted inferences based on the input.

\subsubsection{Mamba}
To address the issues presented in \cite{s4}, Mamba \cite{gu2023mamba} introduces a selective scanning mechanism, where the parameters \(B\in\mathbb{R}^{{B}\times{L}\times{ N}}\) , \(C\in\mathbb{R}^{{B}\times{L}\times{ N}}\) and \(\Delta\in\mathbb{R}^{{B}\times{L}\times{ D}}\) are derived from \(x(t)\in\mathbb{R}^{B \times L \times D} \). This enables Mamba to capture contextual information and dynamically adjust weights. But, the existing Mamba framework has a unidirectional aggregation property. Thus, in this paper, we propose a recurrent representation to capture the contextual information of the spectral data. 
    
\subsection{The Overall Architecture}
Mathematically, let \(X\in\mathbb{R} ^{{H}\times{W}\times {S}} \) denote the 3D volume of a pathology MHSI,  where \({H} \times{W} \) is the spatial resolution, and \({S}\)  is the number of spectral bands. The goal of image segmentation is to predict the per-pixel label map \(\hat{Y}\in{\{0, 1\}^{{H}\times{W}}}\). Our training set is \(D = \{(X_i, Y_i)\}^{N}_{i=1}\), where \(Y_i\) denotes the per-pixel groundtruth for MHSI \(X_i\), \(N\) is the number of training set.
\par
As shown in Fig.~2, our dual-stream model is the first to jointly explore remote spatial relationships and global spectral sequence features in medical hyperspectral image segmentation using Mamba. It comprises a spatial stream and a spectral stream. The spectral stream extracts global, low-redundancy spectral features, which are integrated from the channel dimension through concatenation after the second stage of the spatial stream's encoder, to aggregate spatiospectral features. 
\par 
\subsection{Spatial Stream}
Inspired by \cite{ruan2024vmunet}, the spatial stream follows the U-shaped architecture based on Mamba, treating MHSI as 2D data to extract spatial information. As depicted in Fig.~2, the MHSI \(X\) is first processed by a patch embedding layer that divides it into non-overlapping patches and maps the dimensions to \({L}\), resulting in \(X_{spa}^{in}\).  This is then fed into the encoder to extract spatial features, where patch merging in the first three stages reduces the spatial dimensions while increasing the channel dimensions. The decoder follows a similar structure with four stages. Before the last three stages, patch expansion is applied to reduce channel dimensions and increase spatial dimensions. 
\par
The encoder uses [\(n\), \(n\), \(n\), \(n\)] VSS blocks, and the decoder uses [\(n\), \(n\), \(n\), \(n/2\)] VSS blocks, where \(n\) is the number of VSS blocks per stage. Skip connections transfer high-resolution features from the encoder to the decoder. Finally, a final projection layer adjusts the feature size to match the segmentation target.

\subsection{Spectral Stream}
The spectral stream employs a basic Transformer paradigm\cite{Transformer}, but we innovatively introduce learning spectral sequence representations by bi-directional mamba. It comprises three key modules: Depth-wise Convolution (DwConv), Spectral Sequence Representation Layers (SSRL), and Feed Forward Network (FFN), and can be formulated as: \par
\vspace{-10pt}
\begin{equation}
\begin{aligned}
& X_{spe} = DwConv(X_{spe}^{in}), \\
 &X_{spe}^{'} = SSRL(Norm(X_{xpe}))+ Flatten(X_{spe}),\\
 &X_{spe}^{out} = FFN(Norm(X_{xpe}^{'}))+x_{spe}^{'}, 
\end{aligned}
\label{eq:spectral stream}
\end{equation}
where \(X_{spe}^{in}\) and \(X_{spe}^{out}\) are the input and output of the spectral stream, while \(X_{spe}\) and \(X_{spe}^{'}\) represent the feature after \(DwConv(*)\) and \(SSRL(*)\), respectively. 
\par
We first extend a  spectral feature dimension \( {L}_{spe}\) to MHSI \(X\), integrating \( {L}_{spe}\)  into the batch \({B}\), resulting in \(X_{spe}^{in} \in \mathbb{R}^{{(B\cdot S)} \times {H} \times {W} \times {L}_{spe}}\). Subsequently,  we propose using DwConv to integrate redundant spatial information into the spectral features, thereby avoiding spatial noise interference in spectral feature extraction.  We design the \(SSRL(*)\) operation to represent the feature map \(X_{spe}\) as continuous spectral sequence features. Concretely,  we flatten \(X_{spe}\)  to spectral sequence tokens \(T_{spe} \in \mathbb{R}^{{(B\cdot {H}\cdot{W})}\times {S}\times{L}_{spe}}\)  with \({S}\) spectral tokens. Due to the unidirectional nature of the state space model, a single-directional representation limits global spectral feature capture.  Thus, we propose a recurrent calculation for each token of \(T_{spe} =\{[T_{0}, T_{1},..., T_{{S}}]| T_j\in\mathbb{R}^{1\times {L}_{spe}}\} \) as below:  
\begin{equation}
\begin{aligned}
&  \tilde{h}_{j} = \bar{A}_{spe}\tilde{h}_{j-1} +  \bar{B}_{spe}T_j,\\
 &{P}_{j} = \bar{C}_{spe}\tilde{h}_{j}, 
\end{aligned}
\label{eq:SSRL}
\end{equation}
where \({P}_{j}\) denotes the \(j\)th token of output sequences. \( \bar{A}_{spe}\), \( \bar{B}_{spe}\) and \( \bar{C}_{spe}\) represent the trainable parameters in Spectral Sequence Representation Layers (SSRL). 
\par After \(SSRL(*)\) operation, we can obtain a set of output sequences \(P_{spe} = \{[P_{0}, P_{1},..., P_{{S}}]| P_j\in\mathbb{R}^{1\times {L}_{spe}}\}\). Next, the output \(T_{spe} \in \mathbb{R}^{{B\cdot {H}\cdot{W}}\times {S}\times{L}_{spe}}\) is element-wise added to \(P_{spe}\). 
Finally,  we utilize a Feedforward Neural Network (FFN) to enhance the individual component of spectral tokens.

\section{EXPERIMENTS}
\label{sec:pagestyle}
\subsection{Datasets}
We conduct experiments on the Hyperspectral Cervical Cancer (HCC) dataset with \(954\) scenes and the public Muti-Dimensional Choledoch (MDC) Dataset with \(538\) scenes. 
\par
Both datasets contain high-quality labels for binary MHSI segmentation tasks. The wavelengths from \(450nm\) to \(750nm\) for HCC and \(550nm\) to \(1000nm\) for MDC, resulting in \(40\) and \(60\) spectral bands for each scene. The sizes of single band image in HCC and MDC are resized to \(256\times 256\) and \(256\times 320\), respectively. The HCC dataset is randomly divided into four subsets. Each subset is used as the test set, while the other three are used for training.  For the MDC dataset, following \cite{hypernet, xie2023}, we partition the datasets into training, validation, and test sets using a patient-centric hard split approach with a ratio of 3:1:1. Specifically, we ensure that there is no overlap of data from the same patient across different sets.

\subsection{Experimental Setup}
We employ data augmentation techniques like rotation and flipping, and train with an Adam optimizer using a combination of dice loss and cross-entropy loss, with a batch size of 8 for 100 epochs. Segmentation performance is evaluated using the Dice-Sørensen coefficient (DSC), Hausdorff Distance (HD), computational complexity (MACs, G), and Throughput (images/s). The implementation is based on Pytorch, utilizing four NVIDIA GeForce RTX 3090 GPUs.

\subsection{Comparison with State-of-the-Arts}
Table \ref{Tab:comparison_sota} shows comparisons on MDC dataset and HCC dataset.  The results show that our proposed method achieves state-of-the-art performance with low resource consumption. It can be observed that CNN-based methods generally have faster processing speeds compared to Transformer-based methods, but they are slightly inferior in segmentation performance (DSC \& HD). RNN-based methods, due to their proficiency in handling sequential problems, excel at modeling spectral relationships, leading to decent performance. Our method achieves the best segmentation results in terms of DSC while maintaining a fast speed and low computational resources. 
\par
We assess feature redundancy by computing the Pearson correlation coefficient between different feature channels \cite{wang2022revisiting}. As shown in Table \ref{redundancy}, our method effectively reduces feature redundancy, improving segmentation performance. This improvement is driven by mamba's parameter \(A\), which aggregates the essential features of all spectral information.
\begin{table}[t]
\centering
\vspace{-5pt}
\renewcommand\arraystretch{0.6}
\setlength{\tabcolsep }{13pt}
\caption{Feature redundancy of methods on MDC dataset}
\begin{tabular}{lcc}
\toprule
\textbf{Method} & \textbf{DSC} & \textbf{Redundancy} \\ 
\midrule
\textbf{CNN \cite{yun2023factor}}& 70.88& 0.3699      \\ 
\textbf{Transformer \cite{chen2021transunet}} & 72.85        & 0.2627      \\ 
\emph{\textbf{Ours}}            & 74.50& 0.1743      \\ 
\bottomrule
\end{tabular}
\label{redundancy}
\vspace{-15pt}
\end{table}

\subsection{Ablation Study}
\subsubsection{The effectiveness of spectral stream }
We conduct  an ablation study to evaluate the effectiveness of integrating spectral information into the spatial stream.  As shown in Table \ref{Tab:ablation1},  integrating spectral information improves segmentation performance by at least \(2.74\% \) (\(67.23\) VS. \(69.97\)) in L1. Further improvements are observed when the spectral stream is applied solely in L2 to L4, with L2 showing the largest boost  (\(74.38\)), followed by L3 (\(73.11\)), and L4 (\(72.85\)). These findings suggest that spectral features are most effective at intermediate network layers. Inserting spectral information in L2 enhances  low-level semantic features learning, offering more comprehensive context. Conversely, inserting spectral information at every layer increases computational costs and leads to redundancy, which can cause interference. 
\par
We also assess the model’s ability to distinguish between positive and negative samples using Specificity and Sensitivity.  Although L1 performs slightly better than L2 in terms of Specificity (\(0.9088\) VS. \(0.8660\)),  L2 consistently excels across other metrics. Therefore, we adopt a dual-stream design with the spectral stream integrated at the L2 layer. 

\begin{table}[t]
\vspace{-5pt}
\renewcommand\arraystretch{0.8}
\setlength{\tabcolsep }{6pt}
\centering
\caption{Ablation study (``mean") on the HCC dataset. L1 to L4 represent where spectral stream outputs are inserted into the spatial stream. Best results are\textbf{ highlighted}.}
\begin{tabular}{cccccccc}
\toprule
 \textbf{L1} & \textbf{L2} &\textbf{L3} & \textbf{L4} & \textbf{DSC} & \textbf{HD} & \textbf{Specificity} & \textbf{Sensitivity} \\ 
\midrule
$\times$ & $\times$ & $\times$ & $\times$ & 67.23 & 88.14 & 0.8912 & 0.8260 \\ 
$\checkmark$ & $\times$ & $\times$ & $\times$ & 69.97 & 77.61 & \textbf{0.9088} & 0.7812 \\ 
$\times$ & $\times$ & $\checkmark$ & $\times$ & 73.11 & 78.03 & 0.8839 & 0.8512 \\
$\times$ & $\times$ & $\times$ & $\checkmark$ & 72.85 & 75.55 & 0.8914 & 0.8492 \\ 
$\checkmark$ & $\checkmark$ & $\checkmark$ & $\checkmark$ & 70.22 & 76.84 & 0.8881 & 0.8352 \\ 
$\times$ & $\checkmark$ & $\times$ & $\times$ & \textbf{74.38} & \textbf{74.53} & 0.8934 & \textbf{0.8660} \\ 
\bottomrule     
\end{tabular}
\label{Tab:ablation1}
\vspace{-5pt}
\end{table}

\begin{table}[t]
\renewcommand\arraystretch{0.8}
\setlength{\tabcolsep }{4pt}
\centering
\caption{Ablation Study (in ``mean") on HCC dataset. Tr, SMD\cite{yun2023factor} and Conv mean we replace the SSRL in the spectral stream with Muti-Head Attention,  Spectral Matrix Decomposition and convolution. Best results are \textbf{highlighted}.}
\begin{tabular}{ccccccc}
\toprule
 \textbf{Spectral stream} & \textbf{DSC} & \textbf{HD}  & \textbf{Specificity}  & \textbf{Sensitivity}  & \textbf{Throughput}\\ 
\midrule
{SMD}  &   {71.24} & {83.33} & {0.8823}& {0.8536} & {21.82}\\ 
Tr  &   {71.09} & {77.00} & {0.8941} &{0.8458} & {19.40} \\ 
Conv  &   {70.80} & {77.10} & {0.8819} &{0.8434 } & {13.79} \\ 
mamba  & {69.16} & {83.16} & {0.8871} &{0.8213} & {\textbf{30.07}} \\ 
SSRL  & \textbf{74.38} & \textbf{74.53} & \textbf{0.8944}& \textbf{0.8660} & {25.62} \\ 
\bottomrule    
\end{tabular}

\label{ablation2}
\vspace{-15pt}
\end{table}

\par
\subsubsection{The effectiveness of the SSRL}
We conduct  an ablation study to demonstrate the effectiveness of SSRL's recurrent calculation. Replacing SSRL with Mamba results in a \(5.22\% \) decrease in DSC (\(74.38\) VS. \(69.16\)), confirming that Mamba’s unidirectional information flow limits its ability to fully capture contextual information across spectral bands, reducing spectral utilization. Replacing SSRL with Transformer and SMD led to DSC drops of  \(3.29\% \) (\(74.38\) VS. \(71.09\)) and \(3.14\% \) (\(74.38\) VS. \(71.24\)) respectively, likely due to Transformers being harder to optimize on small datasets and SMD causing spectral information loss. SSRL also outperforms other models in Specificity (0.8934) and Sensitivity (0.8660), accurately identifying both negative and positive samples. Moreover, SSRL shows faster inference speed. While Mamba’s unidirectional computation is expected to excel in Throughput, SSRL still delivers better overall performance.  

\section{CONCLUSION}
This paper introduces a Mamba-Driven Dualstream Network (MDN) for efficient MHSI segmentation. Utilizing Mamba's global context modeling, MDN captures long-range spatial relationships and global spectral features, achieving high segmentation accuracy and fast inference. By integrating spatial information into the spectral dimension and designing a recurrent spectral sequence to capture low-redundancy global features, MDN minimizes spectral redundancy. Evaluations on two MHSI datasets demonstrate that MDN outperforms state-of-the-art methods, offering better segmentation, faster inference, and lower resource consumption.  
\bibliographystyle{splncs04}
\bibliography{refs}

\end{document}